\documentclass[twocolumn,prl,showpacs,color,superscriptaddress,epsfig]{revtex4}
\usepackage[french]{babel}
\usepackage[latin1] {inputenc}
\usepackage{epsfig}

\begin{document}
\title{Magnetic anisotropy of Co$^{2+}$ as signature of intrinsic ferromagnetism in ZnO$:$Co }

\author{P. Sati}
\affiliation{Laboratoire Mat{\'e}riaux et Micro{\'e}lectronique de
Provence, associé au CNRS, Case-142, Université d'Aix-Marseille
III, 13397 Marseille Cedex 20, France}
\author{R.Hayn}
\affiliation{Laboratoire Mat{\'e}riaux et Micro{\'e}lectronique de
Provence, associé au CNRS, Case-142, Université d'Aix-Marseille
III,   13397 Marseille Cedex 20, France}

\author{R. Kuzian}
\affiliation{  Institute for Materials Science, Krzhizhanovskogo
3, 03180, Kiev,
 Ukraine}
\author{S. Régnier}
\affiliation{Laboratoire Mat{\'e}riaux et Micro{\'e}lectronique de
Provence, associé au CNRS, Case-142, Université d'Aix-Marseille
III,   13397 Marseille Cedex 20, France}

\author{S.Sch\"afer}
\affiliation{Laboratoire Mat{\'e}riaux et Micro{\'e}lectronique de
Provence, associé au CNRS, Case-142, Université d'Aix-Marseille
III,   13397 Marseille Cedex 20, France}
\author{A.Stepanov}
\affiliation{Laboratoire Mat{\'e}riaux et Micro{\'e}lectronique de
Provence, associé au CNRS, Case-142, Université d'Aix-Marseille
III,   13397 Marseille Cedex 20, France}

\author{C. Morhain}
\affiliation{ Centre de Recherche sur l'Hétéro-Epitaxie et ses
Applications-CNRS, 06560, Valbonne Sophia-Antipolis, France }
\author{C. Deparis}
\affiliation{ Centre de Recherche sur l'Hétéro-Epitaxie et ses
Applications-CNRS, 06560, Valbonne Sophia-Antipolis, France }

\author{M. Laügt}
\affiliation{ Centre de Recherche sur l'Hétéro-Epitaxie et ses
Applications-CNRS, 06560, Valbonne Sophia-Antipolis, France }
\author{M. Goiran}
\affiliation{ Laboratoire National des Champs Magnétiques Pulsés,
 31432, Toulouse, France }

\author{Z. Golacki}
\affiliation{ Institute of Physics, Polish Academy of Sciences,
Al. Lotnikow 32/46, 02-668, Warsaw, Poland }

\begin{abstract}

We report on the magnetic properties of thoroughly-characterized
Zn$_{1-x}$Co$_{x}$O epitaxial thin films, with low Co
concentration, $x=0.003-0.005$. Magnetic and EPR measurements,
combined with crystal field theory, reveal that isolated Co$^{2+}$
ions in ZnO possess  a strong single ion anisotropy which leads
to an "easy plane" ferromagnetic state when the ferromagnetic
Co-Co interaction is considered. We suggest that the peculiarities
of the magnetization process of this state can be viewed as a
signature of intrinsic ferromagnetism in ZnO$:$Co materials.

\end{abstract}

\maketitle

Spintronics, an emerging branch of micro- and nanoelectronics
which manipulates the electron spin rather than its charge, has
need for spin polarization components. In most spintronic devices,
{\it metallic} ferromagnetic  (FM) materials are used to this end.
However the physics of metal-semiconductor injection is
incompatible with the concept of  semiconductor devices,
preventing their application \cite{review}. A suitable solution
would be a FM semiconductor at room-temperature.

The magnetic properties of diluted magnetic semiconductors are due
to the substitution of cations by transition-metal (TM) ions, and
have been extensively studied for at least five decades
\cite{furdyna}. Co-doped ZnO -a possible candidate for high-$T_c$
FM semiconductors - has attracted much interest from both
theoretical and experimental points of view. Yet, there is an
ongoing debate about its magnetic properties. Early theoretical
studies using the  local spin density approximation (LSDA) for
Zn$_{1-x}$Co$_{x}$O found it to be a FM semimetal \cite{sato1}.
Contrary to this, more recent LSDA calculations \cite{risbud,lee}
on large supercells detected a competition between FM and
antiferromagnetic (AFM) interactions, {\sl i.e.} an AFM or
spin-glass
groundstate. 

Experimentally,  high-$T_c$ FM phases in Zn$_{1-x}$Co$_{x}$O
($x=0.1$-$0.25$) were found in thin films produced by pulsed laser
deposition \cite{prellier}, by the sol-gel method \cite{hyeon},
and by rf magnetron co-sputtering \cite{lim}. They were also found
in bulk single crystals prepared by implantation \cite{norton}.
Controversially, {\it  AFM correlations between TM ions and the
absence of any FM bulk phases} were observed in
Zn$_{1-x}$Co$_{x}$O ($x=0.005$-$0.15, 0.2$) samples fabricated by
precursor decomposition \cite{risbud}, in polycrystalline powder
samples \cite{yoon} as well as in thin films \cite{kim2}.

In this rather contradictory situation a major question which
arises is whether a reliable identification of an intrinsic FM
phase of ZnO doped by Co is possible at all.

Here we  address this question  on both experimental and
theoretical grounds. We argue that such an identification requires
a thorough examination of the magnetic properties of Co$^{2+}$
ions in the ZnO lattice, and in particular the magnetic anisotropy
of cobalt. By EPR and magnetic measurements, we first prove that
Co$^{2+}$, which has a spin $S=3/2$, shows a huge single ion
anisotropy of $D{S_{z}}^2$ type, with $D=2.76\,{\rm cm}^{-1}$. We
then validate this result  theoretically by combining crystal
field  theory with an estimate of the crystal field parameters.
Theory and experiment clearly demonstrate that Co substitutes Zn
in our samples. Finally, using a simple model, we show that a FM
ZnO$:$Co would be an "easy plane" ferromagnet exhibiting a
peculiar magnetization process which offers a simple and reliable
way to identify the intrinsic FM phase of ZnO$:$Co.

We focus on the magnetic anisotropy of isolated cobalt in ZnO and
present details of the magnetic properties of epitaxial thin films
with  very low Co concentration varying from x=0.003 to x=0.005
\cite{remarque}. The 1$\mu$m thick samples were grown on sapphire
substrate by plasma-assisted MBE. 2D growth is achieved for growth
temperature of 560°C, i.e. 50°C higher than the optimal growth
temperature used for ZnO, resulting in streaky RHEED patterns. For
this range of Co composition, the rocking curve FWHMs are in the
range of $\omega \sim$ 0.15° along (002), (-105) and (105). The
low $\omega$ values measured both for (-105) and (105), as well as
their similarity, indicate a large column diameter, close to
1$\mu$m. The c-axis of the wurtzite structure is perpendicular to
the film plane. The conductivity of the films is $n$-type, with
residual carrier concentrations $n_{e}$ $<$ 10$^{18}$ cm$^{-3}$, a
doping level well below the Mott transition.

Neglecting the hyperfine interaction, the $^4$A$_2$ ground-state
of Co$^{2+}$ at a tetrahedral site of the ZnO host lattice is
described by the following $S=3/2$ spin Hamiltonian \cite{abragam}
\begin{equation}\label{hspin}
\widehat{H}_{spin}=\mu_{B}g_{\parallel}H_{z}S_{z}+
\mu_{B}g_{\perp}(H_{x}S_{x}+H_{y}S_{y})+DS_{z}^{2} \;.
\end{equation}
The magnetic state of Co$^{2+}$ can thus be parameterized by only
three constants: the two $g$-factors, $g_{\parallel}$ ($H\parallel
c$) and $g_{\perp }$ ($H \perp c$), and the zero-field splitting
constant $D$. Henceforth, we will use (\ref{hspin}) with the
values inferred from our experiments (see below), namely
$g_{\parallel}=2.236$, $g_{\perp}=2.277$ and $D=2.76\,{\rm
cm}^{-1}$, to compute (i) the magnetization and the magnetic
susceptibility of ZnO$:$Co in a simple statistical model of an
ensemble of independent spins; (ii) the angular and field
dependences of the EPR spectra; (iii) the magnetization of small
FM clusters.

\begin{figure}[h]
    \begin{center}
      \includegraphics[bb=20 250 700 750 width=.2,height=.22\textheight]{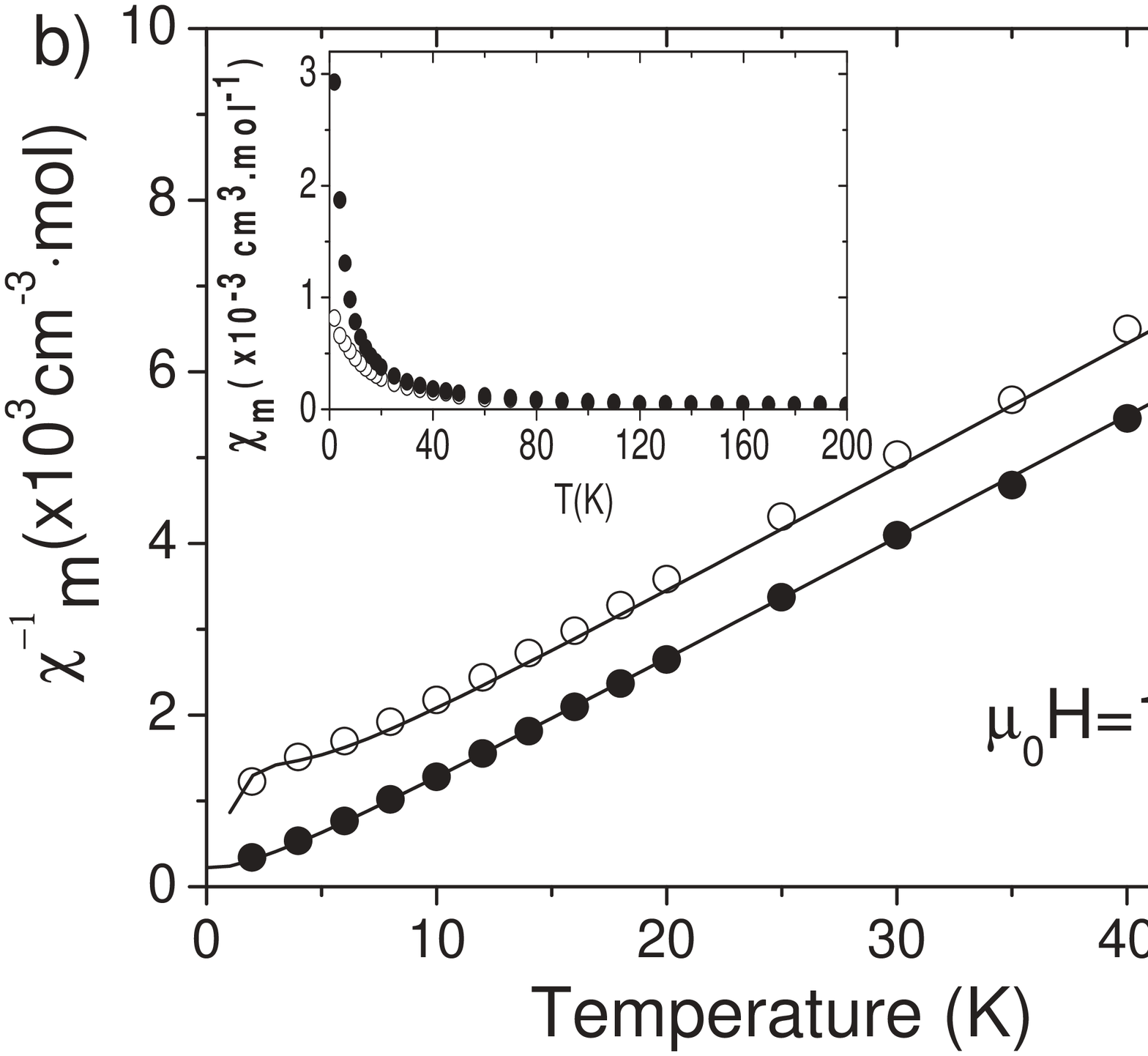}
        \includegraphics[bb=20 250 700 750 width=.2,height=.22\textheight]{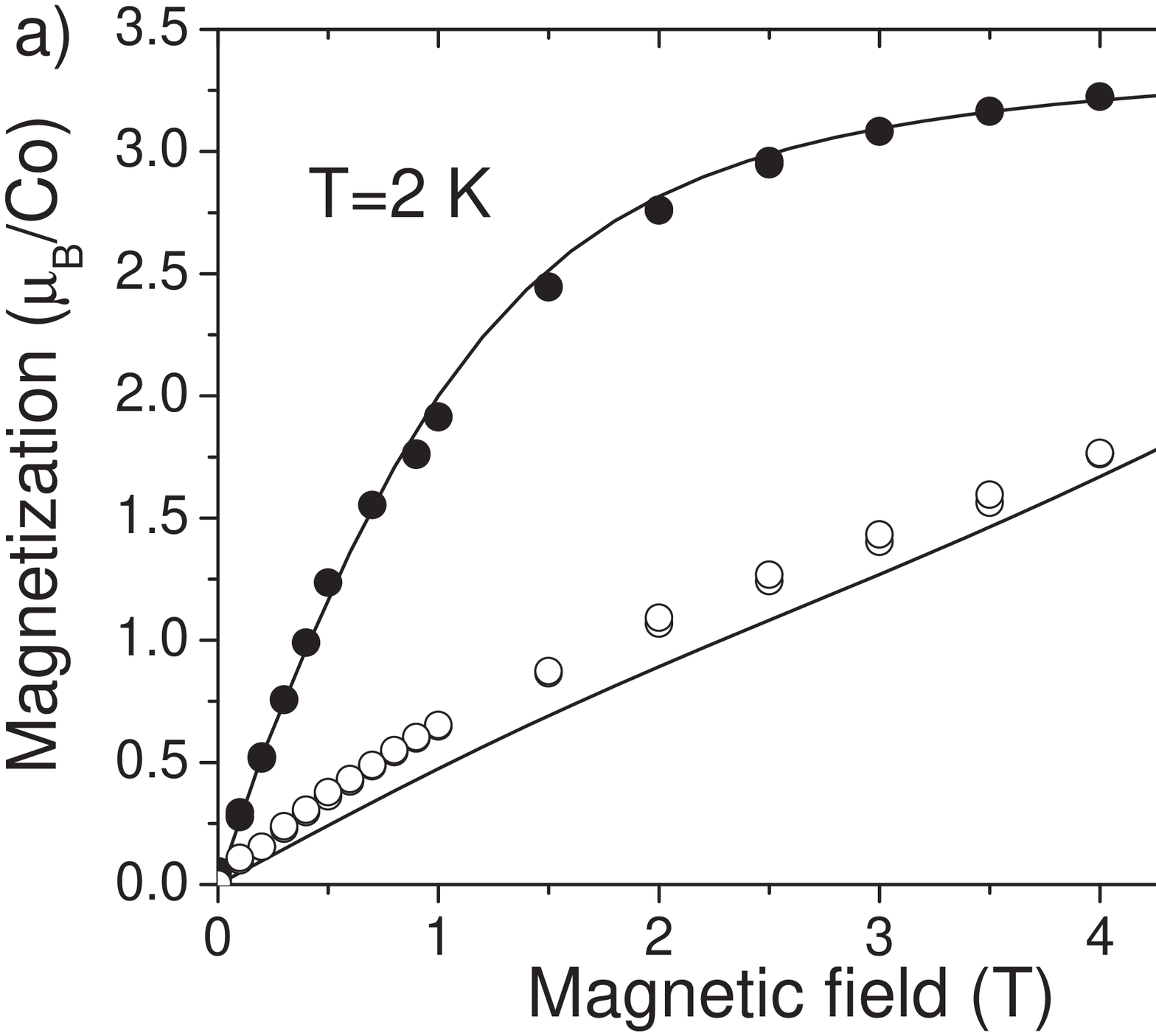}

      \caption{a) Field dependence of magnetization for a Zn$_{x}$Co$_{1-x}$O
      sample with x=0.0028 at T=2K.
b) Inverse of the magnetic susceptibility for the same sample $vs$
temperature  at $\mu_0$H=1 T; the inset shows $\chi (T)$. Full and
open circles are experimental data for  $H \perp c$ and $H
\parallel c$, respectively. Solid lines are  computed according
 to the model discussed in the text.}

\end{center}
\end{figure}

First, we discuss the results of the magnetic measurements of thin
ZnO$:$Co films.  The substrate was cut into thin rectangular
platelets of 3 by $3\,{\rm mm^2}$. In each experiment, up to 10
platelets were piled up to increase the paramagnetic component of
the signal. In addition, undoped ZnO films, deposited on the same
sapphire substrate, were examined.  The latter films served as a
reference of the diamagnetic contribution of the measured signal.
Measurements were performed using a Quantum Design MPMS XL
magnetometer between 300K and 2K in magnetic fields up to 5 T. The
field dependence of the magnetization taken at 2K and the
temperature dependence of the inverse of the  magnetic
susceptibility measured at 1 T for a Zn$_{x}$Co$_{1-x}$O sample
with $x=0.0028$, for two orientations of the magnetic field,
$H\perp c$ and $H\parallel c$, are shown in Figs.~1a and 1b,
respectively.

As is clear from Fig.~1, both the magnetization and the
susceptibility curves reveal a significant magnetic anisotropy.
This anisotropy can be referred to as being of an "easy plane"
type; {\sl i.e.} for a given magnetic field $H$, the magnetization
$M\perp c$ is greater than $M\parallel c$. As expected, the
susceptibility curves show a paramagnetic behavior. Their
deviation from a Curie law finds a natural explanation in a
$S=3/2$ model with a single ion anisotropy, where the only
adjustable parameter is the concentration of Co$^{2+}$ ions.

\begin{figure}[h]
   \begin{center}
      \includegraphics[bb=10 250 700 750 width=.17,height=.23\textheight]{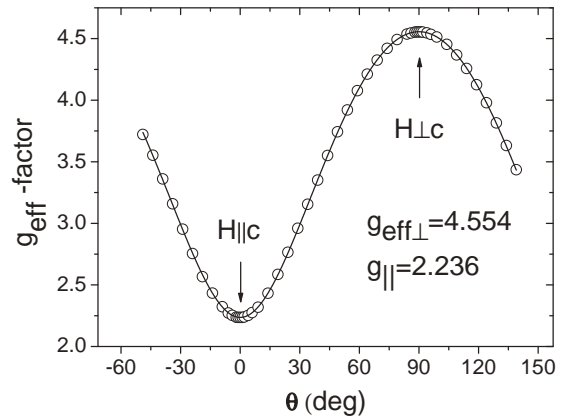}
            \caption{ Angular dependence of the effective g-factor of
      Co$^{2+}$. $\theta =0$ corresponds to $H\parallel c$. The
      open circles are the experimental data; the solid line is a fit
      based on Eq.\ref{hspin}.}

 \end{center}
\end{figure}

We now turn to the results of low-frequency EPR. An EMX Bruker
spectrometer was used to collect spectra in the X-band ($\nu=9.4$
GHz) and in the temperature range 4$-$300 K. A single line with a
partly resolved hf structure for $H\parallel c$ was observed below
$\sim 100$ K. This clearly indicates that $D\gg h\nu$.  As the
temperature is lowered to 4 K, the line intensity increases
monotonically, and roughly follows a simple Curie law, $\sim1/T$,
indicating that the observed EPR signal is due to the low-lying
doublet S$_z=\pm 1/2$ of a S$=3/2$ ground-state manifold
\cite{estle}. Note that the latter observation also allows us to
determine the sign of $D$, $D>0$.

In Fig.~2 the angular dependence of the apparent (effective)
$g$-factor of Co$^{2+}$ in the ZnO lattice is shown, $\theta$
being the angle between the c-axis and the applied field $H$. The
extracted values, $g_{\parallel}=2.236$ and $g_{{\rm eff} \perp
}=4.554$, are very close to those obtained previously for single
crystals \cite{estle} and thin films \cite{jedrecy} of ZnO$:$Co.
The measured $g_{{\rm eff} \perp}$ can be assigned to the lowest
Co$^{2+}$ doublet, yielding $g_{{\rm eff} \perp }\simeq
2g_{\perp}(1-(3/64)(h\nu/D)^2)$ for $D\gg h\nu$ \cite{holton} ,
which with reasonable accuracy reduces to $g_{ \perp}= g_{{\rm
eff}\perp}/2$. The above results also suggest that the Zeeman part
of the Co$^{2+}$ spin Hamiltonian is practically isotropic
($g_{\parallel}\simeq g_{\perp }$), and hence cannot be
responsible for the magnetic anisotropy of ZnO$:$Co. We therefore
need more information about the zero-field splitting constant $D$
which, according to the temperature dependence of the X-band EPR
signal, can be estimated to $2D=5.5 \pm 0.3\,{\rm cm}^{-1}$
\cite{estle}.
\begin{figure}[h]
    \begin{center}
      \includegraphics[bb=10 300 600 750 width=.35,height=.27\textheight]{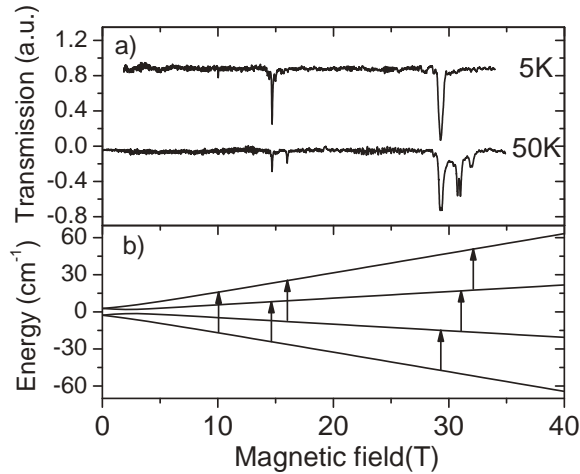}
          \caption{ a) Transmission EPR spectra of ZnO$:$Co taken at
 $\lambda=305 \mu m$ and $\theta=63°$. b)  Energy-level splitting of Co$^{2+}$
in magnetic field at $\theta=63°$ computed according to
Eq.\ref{hspin}. The vertical arrows indicate the observed EPR
transitions.}
  \end{center}
\end{figure}

We have undertaken high-frequency EPR measurements on
Zn$_{1-x}$Co$_{x}$O ($x=0.01-0.02$) single crystals in a large
range of wavelengths $\lambda=3{\rm mm}-0.3{\rm mm}$. Magnetic
fields up to 35 T were provided by the pulsed field facility in
Toulouse. Representative transmission EPR spectra of ZnO$:$Co
taken at the shortest wavelength ($\lambda=305 \mu{\rm m}$) are
displayed in Fig.~3, together with the computed energy level
splitting. The observation of the "forbidden" transitions with
$\Delta m=\pm2$ and $\Delta m=\pm3$ confirms the presence of a
large zero-field splitting term in the spin Hamiltonian. Most
importantly, these experiments allowed us {\it to measure directly
the zero-field splitting constant, $D=2.76 \pm0.01 {\rm
cm}^{-1}$}.

To understand the origin of the magnetic anisotropy of Co$^{2+}$,
we use and extend the standard crystal field (CF) theory
\cite{MacFarlane67,abragam} and estimate the CF parameters. The
Co$^{2+}$ ion is in the $d^{7}$ configuration and Hund's rule
coupling is sufficiently large to restrict the states essentially
to $S=3/2$, {\sl i.e.} the $^4F$ and $^4P$ states. These states
split due to the CF potential $\widehat{H}_{\rm CF}$ and interact
by spin-orbit coupling $\widehat{H}_{\rm SO}=\lambda
\vec{L}\cdot\vec{S}$:
\begin{equation}\label{eq1}
    \widehat{H}=\widehat{H}_{\rm Coul}+\widehat{H}_{\rm CF}+\widehat{H}_{\rm SO}
\quad \quad
    \widehat{H}_{\rm CF}=\widehat{H}_{\rm cub}+\widehat{H}_{\rm trig}
\end{equation}
The $^4P$ states are $15B$ higher in energy ($B$ being the Racah
parameter). The trigonal symmetry of the CF requires three
parameters which, as in Refs.~\cite{MacFarlane67} and
\cite{Koidl}, we denote $\Delta$, $\upsilon$, and $\upsilon'$.

The importance of the non-diagonal matrix element $\upsilon'$ was
first pointed out by MacFarlane \cite{MacFarlane67} but was not
thoroughly treated in  standard text books \cite{abragam}.
MacFarlane's formula \cite{MacFarlane67} is a good approximation
for strong cubic crystal fields ($\Delta \gg 15 B$, as for
Cr$^{3+}$ complexes) but is not applicable in the present case
where $15 B \gg \Delta \gg \upsilon, \upsilon', \lambda$. Here, in
order to calculate $g_{\parallel}$, $g_{\perp}$ and $D$, we derive
a perturbative formula for the parameters of the effective
Hamiltonian ($g_s=2.002$, $k$: reduction factor)
\begin{eqnarray}
\label{pert}
D &=& \frac{\lambda^{2}}{\Delta^{2}}\left[ 2 \upsilon -
\frac{10\sqrt{2}}{3} \upsilon' \left(
1+\frac{4}{75}\frac{\Delta}{B}\right) \right]
\nonumber \\
g_{\parallel} &=& g_s-\frac{8\lambda k}{\Delta} \left[ 1 -
\frac{\upsilon}{3\Delta}+\frac{5\sqrt{2}}{9\Delta}\upsilon'
\left(1+\frac{4}{75}\frac{\Delta}{B} \right)\right]
 \\
g_{\perp} &=& g_s-\frac{8\lambda k}{\Delta} \left[ 1 +
\frac{\upsilon}{6\Delta}-\frac{5\sqrt{2}}{18\Delta}\upsilon'
\left(1+\frac{4}{75}\frac{\Delta}{B} \right) \right] \; .
\nonumber
\end{eqnarray}
Eq.~(\ref{pert}) is more concise and hence more practicable than
its counterpart derived in \cite{mao}.

Numerical diagonalization of the Hamiltonian (\ref{eq1}) within
the subspace of the $^4F$ and $^4P$ states (a $40 \times 40$
matrix) yields the results shown in Table \ref{tab1}. For the CF
parameters obtained from optical measurements \cite{Koidl}, the
anisotropy and gyromagnetic factors agree very well with our
measured values. For the same parameters the perturbative results
of Eq.~(\ref{pert}) are $g_{\parallel}=2.27$, $g_{\perp}=2.30$,
and $2D=4.34\,{\rm cm}^{-1}$, and thus in good agreement with the
numerical diagonalization.

\begin{table}[htbp]
\begin{tabular}{|c|c|c|c|}
  \hline
  & EPR & optics
 &   Harrison \\
\hline $\Delta$ & &4000 &  4000 \\
 $\upsilon$  & & -120 &  53 \\
 $\upsilon'$ & & -320 &  -210 \\
\hline
$g_{\parallel}$ & 2.236 & 2.24 &  2.21 \\
$g_{\perp}$  & 2.277 & 2.28 &   2.23 \\
$2D$  & 5.52 & 4.04 &   3.14 \\
\hline
\end{tabular}
\caption{Measured EPR data, compared to those calculated from CF
theory with parameters from optics \cite{Koidl} or estimated from
the Harrison approach. In all calculations the Racah parameter
$B=760\,{\rm cm}^{-1}$ and the spin-orbit coupling
$\lambda=-143.3\,{\rm cm}^{-1}$ \cite{Koidl} were used. The energy
unit is inverse centimeters.} \label{tab1}
\end{table}

To obtain more microscopic information about the CF parameters
$\Delta$, $\upsilon$ and $\upsilon'$, we consider the
hybridization contribution to the $d$-level splitting
$E_m=(t_{pdm})^2/\Delta_{pd}$, with the Slater-Koster hopping
parameter $t_{pdm}$ ($m=\sigma$ or $\pi$) parameterized according
to Harrison \cite{harrison,kuzmin}. Adjusting the charge-transfer
energy $\Delta_{pd}$ to the optical cubic splitting $\Delta$ gives
$\Delta_{pd}=3.6$~eV and $k=0.85$ and leads to reasonable values
of $g_{\parallel}$, $g_{\perp}$, and $2D$ (last column of Table
\ref{tab1}). So, we may give to the CF parameters that were
derived from the optical measurements a microscopic foundation.
Also, these parameters are very specific to the tetrahedral
environment of Co with trigonal splitting which suggests that our
experimental data can only be explained if Co replaces Zn.

\begin{figure}[h]
  \includegraphics[bb=10 250 700 750 width=.35,height=.24\textheight]{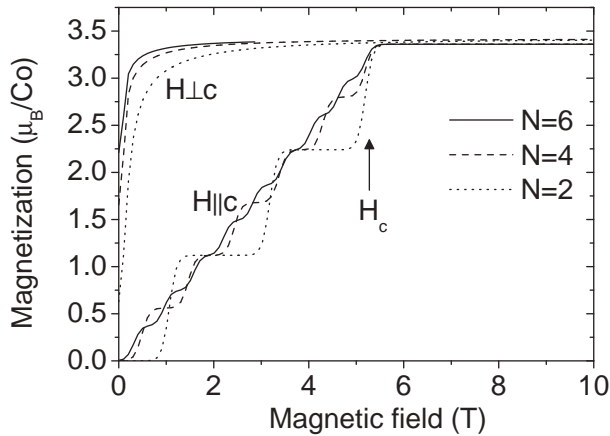}

  \caption{Magnetization of FM Co$^{2+}$ clusters ($N=2, 4,
  6$) as a function of the applied magnetic field, calculated at
  $T=0.1 K$ for $J=-21 {\rm cm}^{-1}$.}  
\end{figure}

We finally discuss the possible magnetic states that would arise
if the individual Co$^{2+}$ ions in the ZnO lattice coupled
ferromagnetically. This can be studied by adding a Heisenberg term
($\sum J S_{i}S_{j}$ with $J<0$) to the Hamiltonian (\ref{hspin})
\cite{brey}. The magnetization curves, obtained by exact numerical
diagonalization of small $S=3/2$ clusters (with $N=2,4,6$), shown
in Fig.~4, are rather insensitive to the exact value of the FM
coupling, provided that $-J\gg D$. As can be seen from the figure,
$M(H)$ strongly depends on the orientation of the magnetic field:
for $H \perp c$, saturation is reached at very low fields; for
$H\parallel c$, by contrast, the magnetization rises at first
essentially linearly (for $N\gg 1$) and saturates only at a
critical field, $H_{c}$. In the limit of $N\gg 1$, we have
$g_\parallel\mu_B H_c = 2D$, and hence $\mu_0 H_c=5.3\,{\rm T}$
for FM ZnO$:$Co. We also note that the saturated magnetization
does not depend on the magnetic field orientation. To our
knowledge the above described  magnetic behavior has not been seen
yet. In contrast to this prediction an "easy axis" magnetic
anisotropy \cite{rode,jpark,dinia} or the absence of any
anisotropy \cite{prellier} were found in ferromagnetic
polycrystalline ZnO:Co thin films, clearly indicating that Co
electronic states other that $3d^7$ are responsible for the
observed ferromagnetic signal.

In summary, we have presented an exhaustive study of the magnetic
properties of Co-doped ZnO thin films in a low-concentration
regime. Our experimental results and theoretical calculations
clearly demonstrate a strong anisotropy of Co$^{2+}$ ions in the
ZnO lattice which affects the magnetic ground-state of ZnO$:$Co
leading to an "easy plane" ferromagnet. In the absence of
consensus regarding the magnetic properties of ZnO$:$Co we argue
that the study of its magnetic anisotropy offers simple criteria,
both experimental and theoretical, for the identification of the
intrinsic ferromagnetism in this material.

Financial support by the NATO science division (grant CLG 98 1255)
is gratefully acknowledged.


\begin{thebibliography}{99}
\bibitem{remarque}During our work on Zn$_{1-x}$Co$_{x}$O  a variety of epitaxial films
with $x=0.003-0.3$
  have been fabricated and
 tested. We found that starting from
 x=0.03-0.05    AFM correlations
 between  Co$^{2+}$  determine the magnetic properties of these compounds.
 Therefore we believe that ZnO$:$Co is
 an intrinsically AFM material even when heavily
 electron doped ($ n_e\geq 3\cdot 10^{20}cm^{-3}$).
\end{thebibliography}
\end{document}